\documentclass[prb,twocolumn,amsmath,showkeys,showpacs,floatfix]{revtex4}
\usepackage{graphicx}
\advance\topmargin by 1 cm
\def\bildA{%%%
\begin{figure}[htb]
\begin{center}
  \includegraphics[width=0.7\columnwidth]{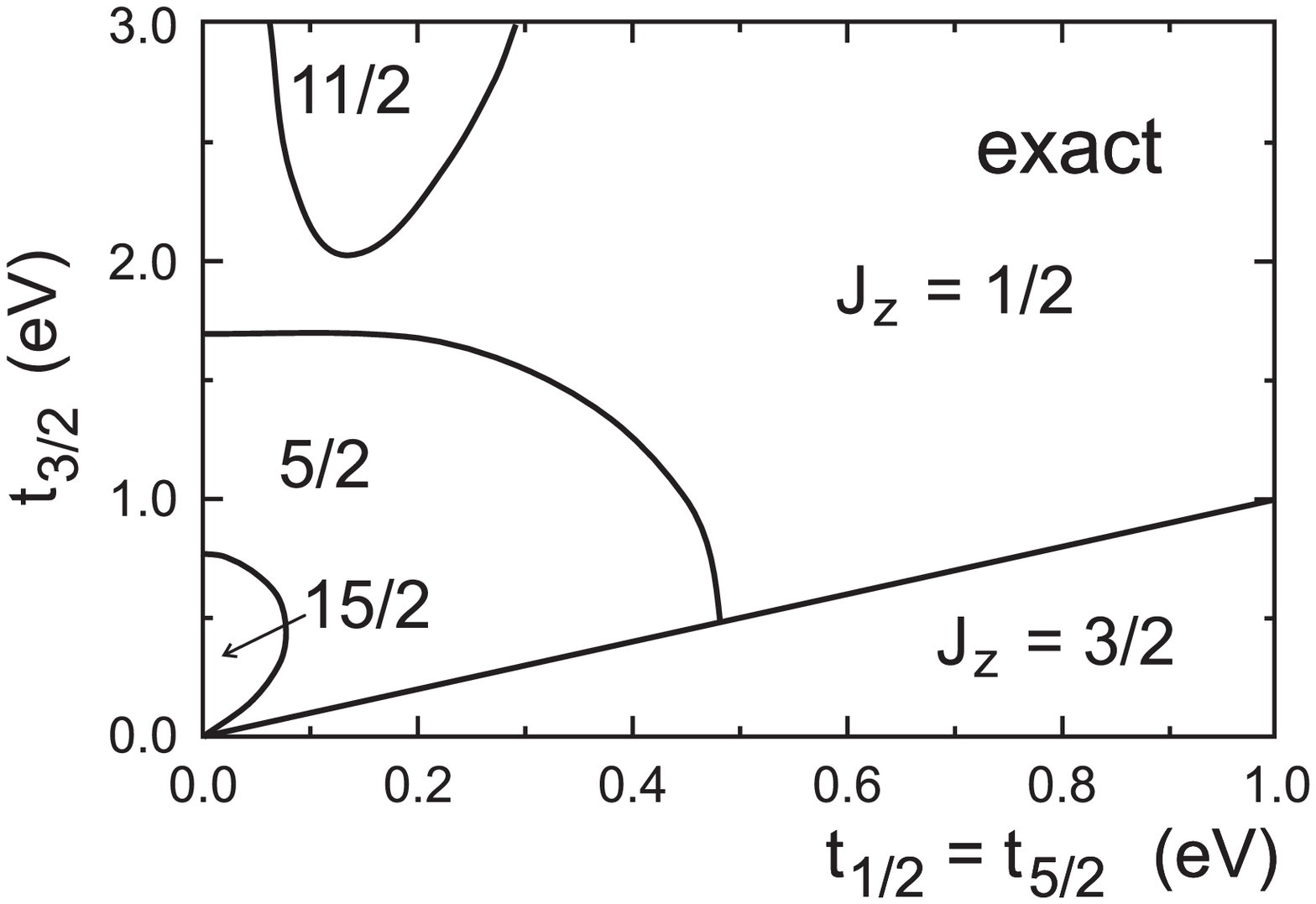}
\end{center}
\caption{Phase diagram for the $f^2$-$f^{3}$ model 
  of Ref.~\protect\onlinecite{I}
  in the $t_{1/2}$\,-\,$t_{3/2}$ plane ($t_{5/2}$\,=\,$t_{1/2}$), 
  derived from the total magnetization ${\cal J}_z$  
  ($z$ component of the total angular momentum) of the ground state
  for infinitesimal field $h_n=0^+$. 
  Numbers indicate  ${\cal J}_z$, which is a good quantum number.
}\label{figCompare}
\end{figure} 
}%%%%%%%%%%%%%%%%%%%%%%%%%%%%%%%%%%%%%%%%%%%%%%%%%%%%%%%%%%%%%%%%%%%%%%%%
\def\bildB{%%%%%%%%%%%%%%%%%%%%%%%%%%%%%%%%%%%%%%%%%%%%%%%%%%%%%%%%%%%%%%
\begin{figure}[htb]
\begin{center}
  \includegraphics[width=0.7\columnwidth]{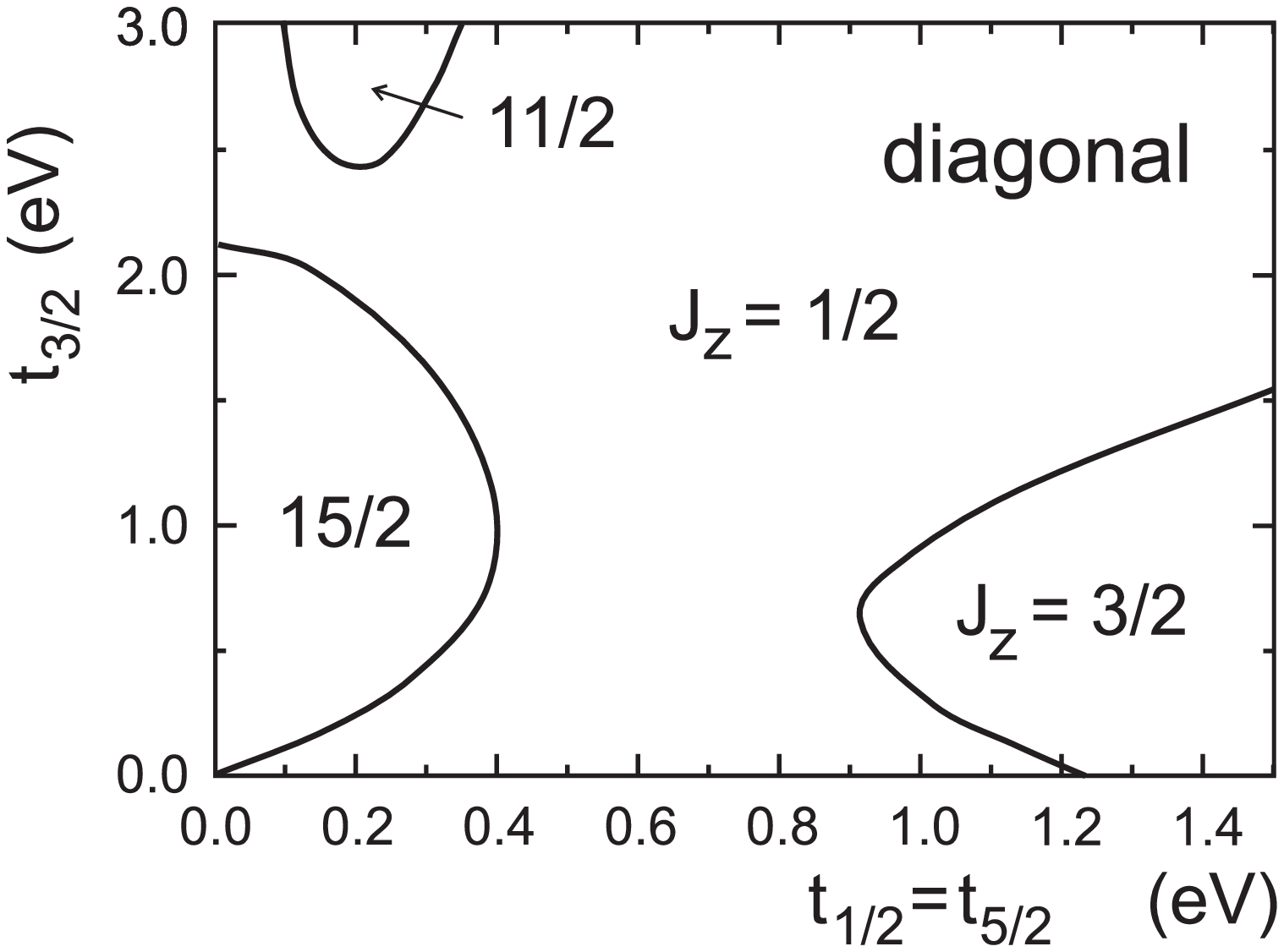}
\end{center}
\caption{Phase diagram  within the diagonal approximation (\ref{eqLDAPlusU2})
  in the $t_{1/2}$\,-\,$t_{3/2}$ plane ($t_{1/2}=t_{5/2}$),
  derived from the total magnetization ${\cal J}_z$
  of the ground state for infinitesimal field $h_n=0^+$.
  In comparison with Fig.~\ref{figCompare}, 
  note the different plot ranges.
}\label{fig:LDA+U}
\end{figure} 
}%%%%%%%%%%%%%%%%%%%%%%%%%%%%%%%%%%%%%%%%%%%%%%%%%%%%%%%%%%%%%%%%%%%%%%%%
\def\bildC{%%%%%%%%%%%%%%%%%%%%%%%%%%%%%%%%%%%%%%%%%%%%%%%%%%%%%%%%%%%%%%
\begin{figure}[h t b]
\begin{center}
  \includegraphics[width=1.0\columnwidth]{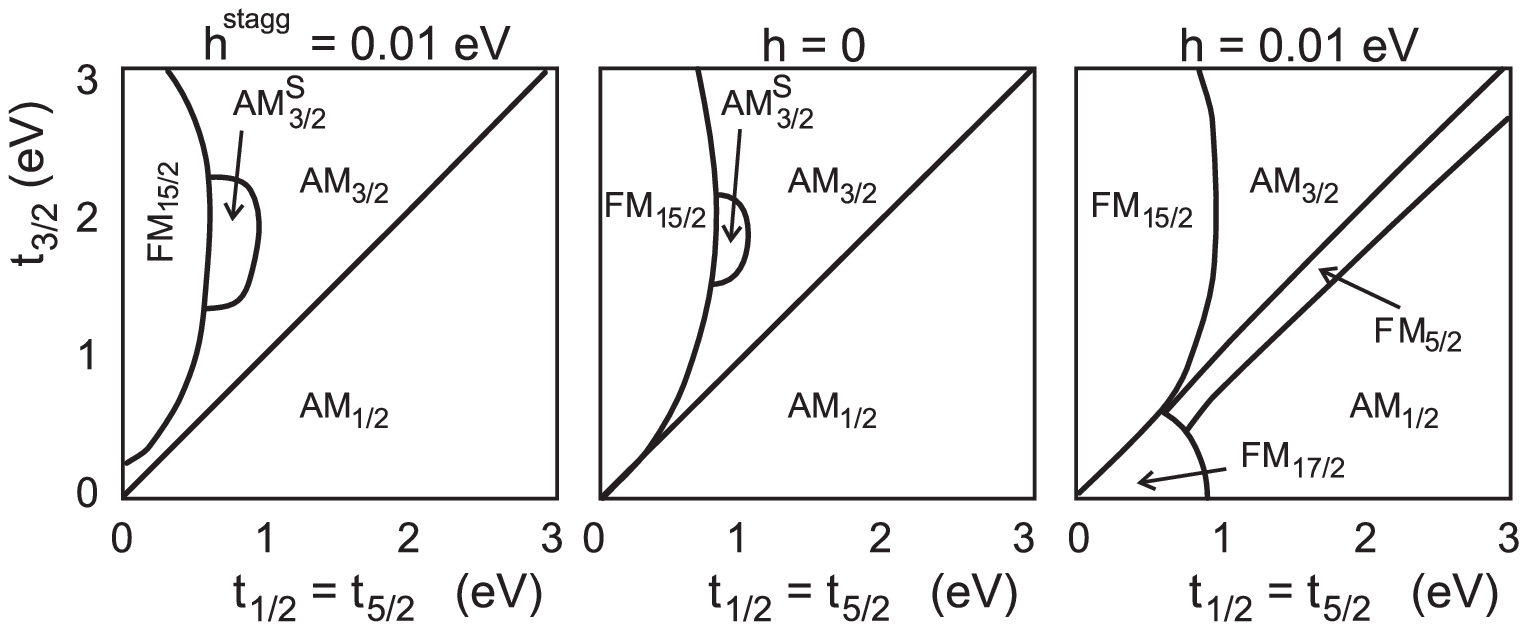}
\end{center}
\caption{
  Phase diagram in the  $t_{1/2}$\,-\,$t_{3/2}$ plane
  (with $t_{1/2}=t_{5/2}$) in the Hartree-Fock approximation
  with ferromagnetic (right), vanishing (middle), 
  and antiferromagnetic (left) fields.
  The phases are chararacterized from the $z$~components 
  of the angular momentum and the state occupation
  and labled according to the notation of 
  Fig.~\ref{fig:spinSchemeMagnet} and \ref{fig:spinSchemeExtra}.  
}\label{figHFMTT}
\end{figure} 
}%%%%%%%%%%%%%%%%%%%%%%%%%%%%%%%%%%%%%%%%%%%%%%%%%%%%%%%%%%%%%%%%%%%%%%%%
\def\bildD{%%%%%%%%%%%%%%%%%%%%%%%%%%%%%%%%%%%%%%%%%%%%%%%%%%%%%%%%%%%%%%
\begin{figure}[h t b]
\begin{center}
  \includegraphics[width=0.85\columnwidth]{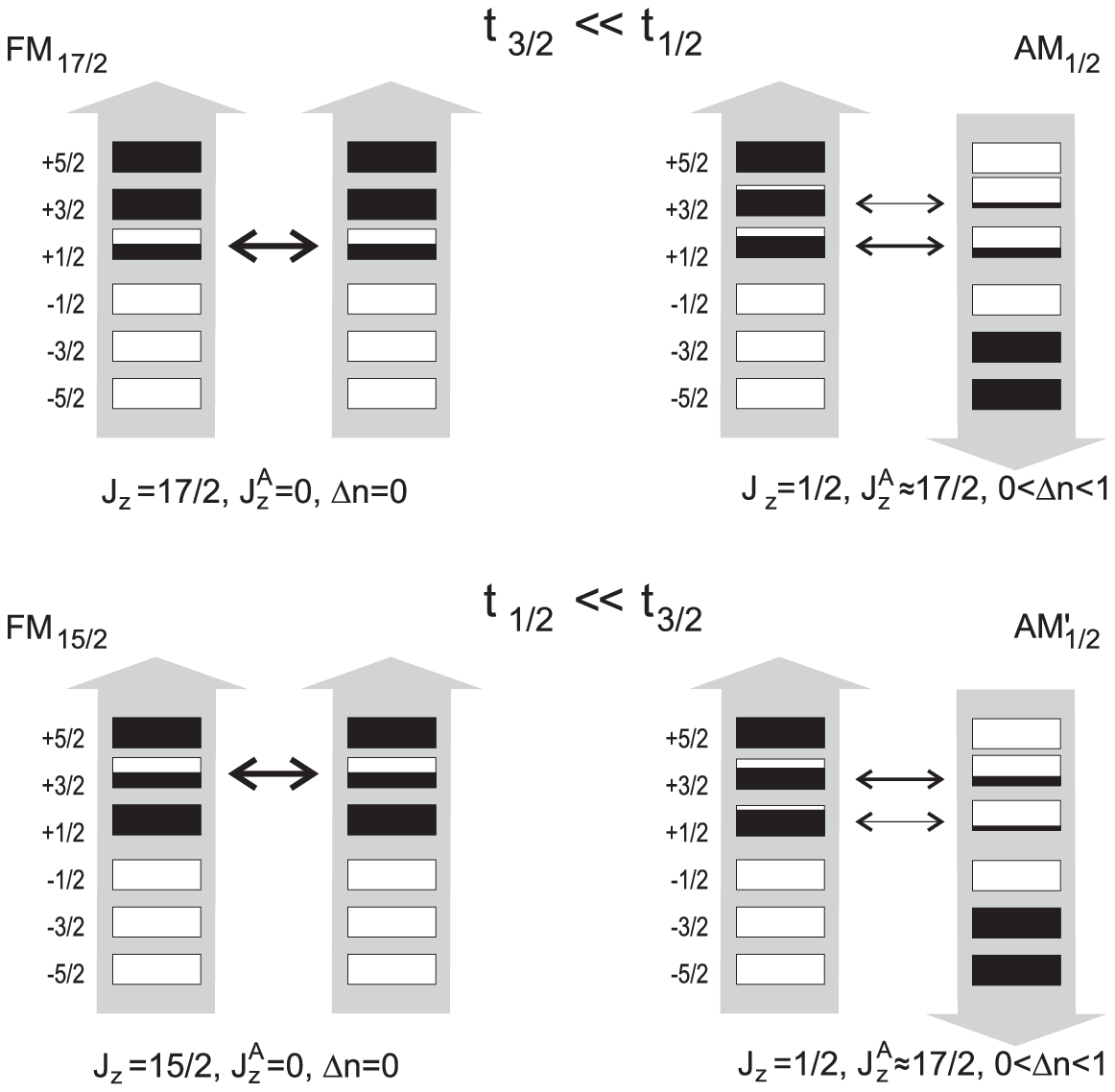}
\end{center}
\caption{
  Configurations relevant for weakly anisotropic hopping (hybridization) in the 
  presence of ferromagnetic (FM, left column) and antiferromagnetic 
  (AM, right column) fields for 
  $ t_{ 3/2} \ll t_{ 1/2}= t_{ 5/2} $ (top row)
  and $  t_{ 1/2}= t_{ 5/2} \ll t_{ 3/2}  $  (bottom row).
  Rectangles symbolize the $j_z$~orbitals on both sites, 
  with black areas presenting their occupation. 
  The thickness of horizontal arrows between orbitals
  symbolizes their contributions to the kinetic energy gain. 
  The arrow-shaped background emphasizes the role of the fields.
  Note the charge disproportionation $0<\Delta n_f<1$ 
  in the antiferromagnetically aligned states. 
}\label{fig:spinSchemeMagnet}
\end{figure}
}%%%%%%%%%%%%%%%%%%%%%%%%%%%%%%%%%%%%%%%%%%%%%%%%%%%%%%%%%%%%%%%%%%%%%%%%
\def\bildE{%%%%%%%%%%%%%%%%%%%%%%%%%%%%%%%%%%%%%%%%%%%%%%%%%%%%%%%%%%%%%%
\begin{figure}[h t b]
\begin{center}
  \includegraphics[width=0.7\columnwidth]{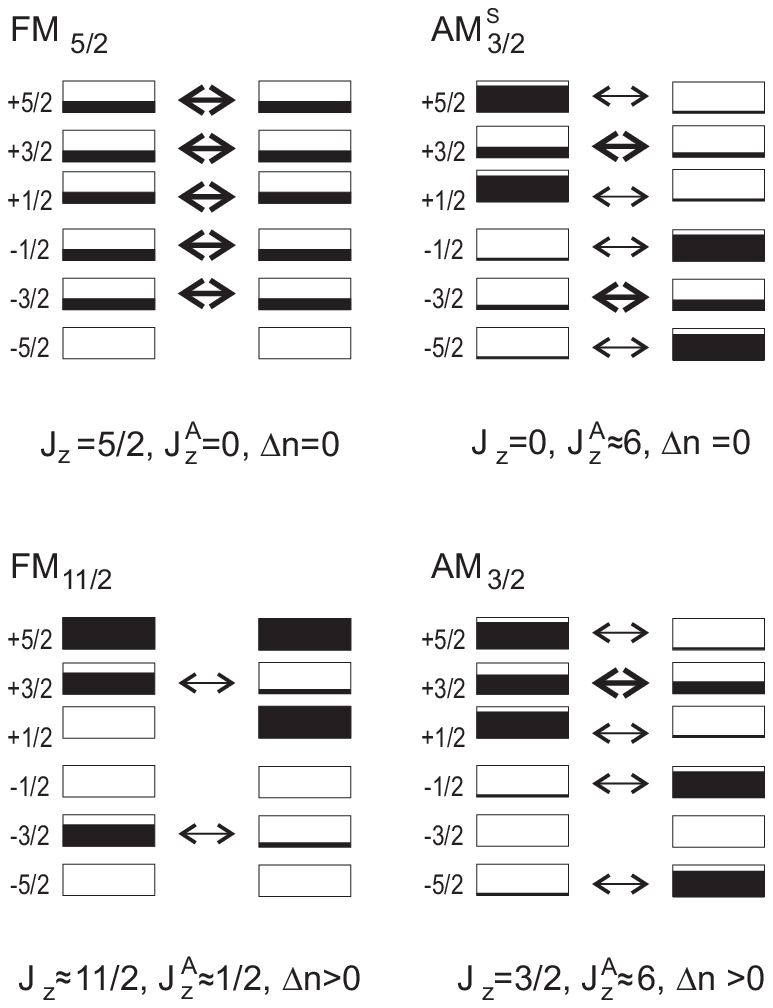}
\end{center}
\caption{
  Configurations relevant for large hopping: 
  The upper left panel (FM$_{5/2}$) shows an overall 
  ferromagnetic arrangement with all electrons 
  contributing to the kinetic energy gain.
  The upper right panel (AM$^{\mathrm{S}}_{3/2}$) presents an overall 
  antiferromagnetic arrangement dominated by $t_{3/2}$ 
  hopping which is symmetric with respect to the site index.
  $t_{3/2}$ hopping dominates also in the lower row.
  In the lower left panel (FM$_{11/2}$), 
  the overall arrangement is of ferromagnetic type,
  whereas in the lower right panel (AM$_{3/2}$) 
  it is of antiferromagnetic character. 
  As in Fig.~\ref{fig:spinSchemeMagnet}, 
  black areas symbolize the orbital occupation
  and the linewidth of the arrows symbolize the corresponding 
  contributions to the kinetic energy gain.
  Field directions are not indicated, in order
  to emphasize the role of the hopping for 
  the formation of these states.
}\label{fig:spinSchemeExtra}
\end{figure}
}%%%%%%%%%%%%%%%%%%%%%%%%%%%%%%%%%%%%%%%%%%%%%%%%%%%%%%%%%%%%%%%%%%%%%%%%
\def\bildF{%%%%%%%%%%%%%%%%%%%%%%%%%%%%%%%%%%%%%%%%%%%%%%%%%%%%%%%%%%%%%%
\begin{figure}[hbt]
\begin{center}
  \includegraphics[width=0.75\columnwidth]{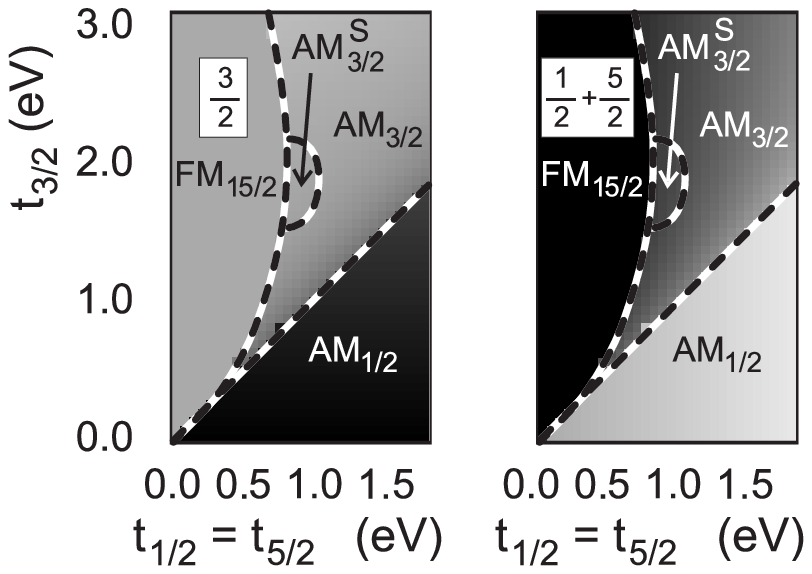}
\end{center}
\caption{
  Ratio (a)  $T_{3/2}/t_{3/2}$ and (b)  $(T_{1/2}+T_{5/2})/t_{1/2}$
  in the Hartree-Fock approximation in the  $t_{1/2}$\,-\,$t_{3/2}$ 
  plane (with $t_{1/2}=t_{5/2}$ and $h_n=0^+$) 
  given as gray-scale  plot (black: 0, white: 1.5). 
  Phase boundaries of the middle panel of Fig.~\ref{figHFMTT}
  are included for orientation. 
}\label{figHFPartial}
\end{figure}
}%%%%%%%%%%%%%%%%%%%%%%%%%%%%%%%%%%%%%%%%%%%%%%%%%%%%%%%%%%%%%%%%%%%%%%%%
\def\bildG{%%%%%%%%%%%%%%%%%%%%%%%%%%%%%%%%%%%%%%%%%%%%%%%%%%%%%%%%%%%%%%
\begin{figure}[h t b]
\begin{center}
  \includegraphics[width=1.0\columnwidth]{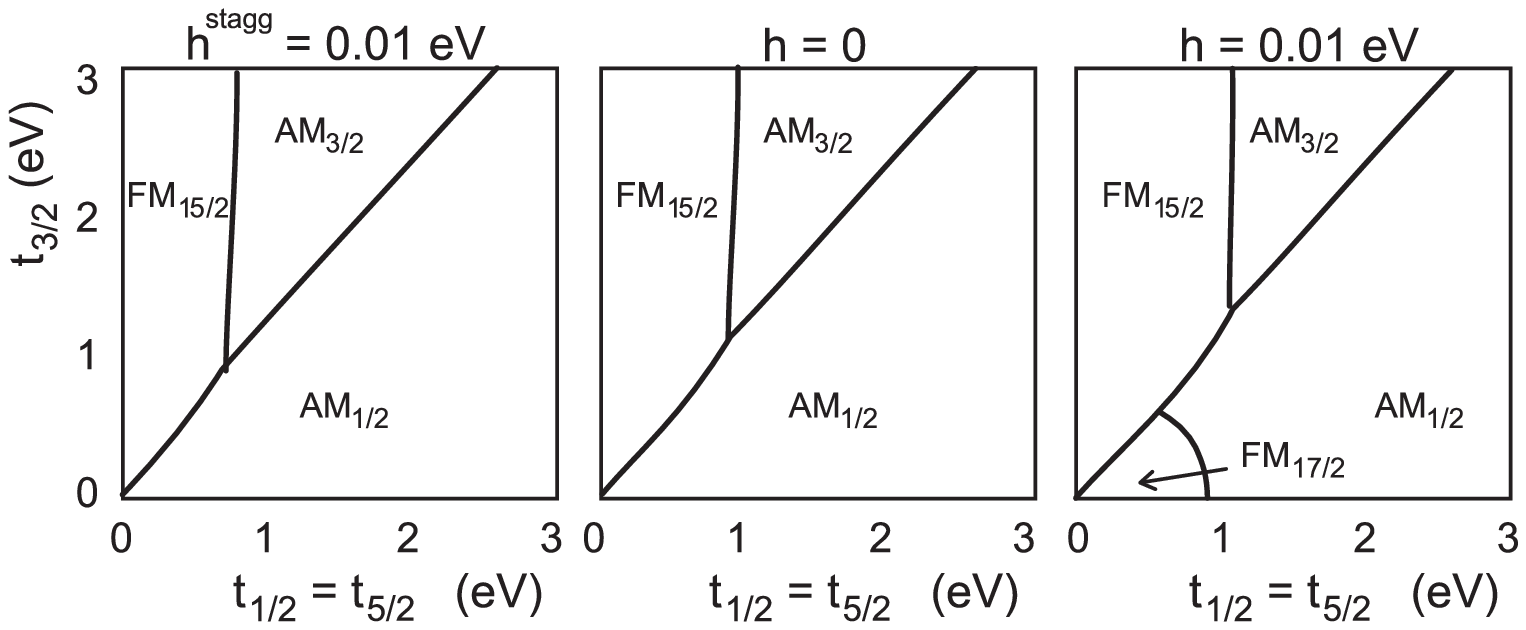}
\end{center}
\caption{
  Hartree-Fock approximation with variational space restricted to
  $j_z$ eigenstates (UHF):
  Phase diagram in the  $t_{1/2}$\,-\,$t_{3/2}$ plane
  ($t_{1/2}=t_{5/2}$) 
  with ferromagnetic (right), vanishing (middle), 
  and antiferromagnetic (left) fields.
  The phases are chararacterized from the $z$~components 
  of the angular momentum and the state occupation, cf. 
  Fig.~\ref{fig:spinSchemeMagnet} and \ref{fig:spinSchemeExtra}.
}\label{figUHF}
\end{figure}
}%%%%%%%%%%%%%%%%%%%%%%%%%%%%%%%%%%%%%%%%%%%%%%%%%%%%%%%%%%%%%%%%%%%%%%%%
\def\bildH{%%%%%%%%%%%%%%%%%%%%%%%%%%%%%%%%%%%%%%%%%%%%%%%%%%%%%%%%%%%%%%
\begin{figure}[hbt]
\begin{center}
  \includegraphics[width=0.85\columnwidth]{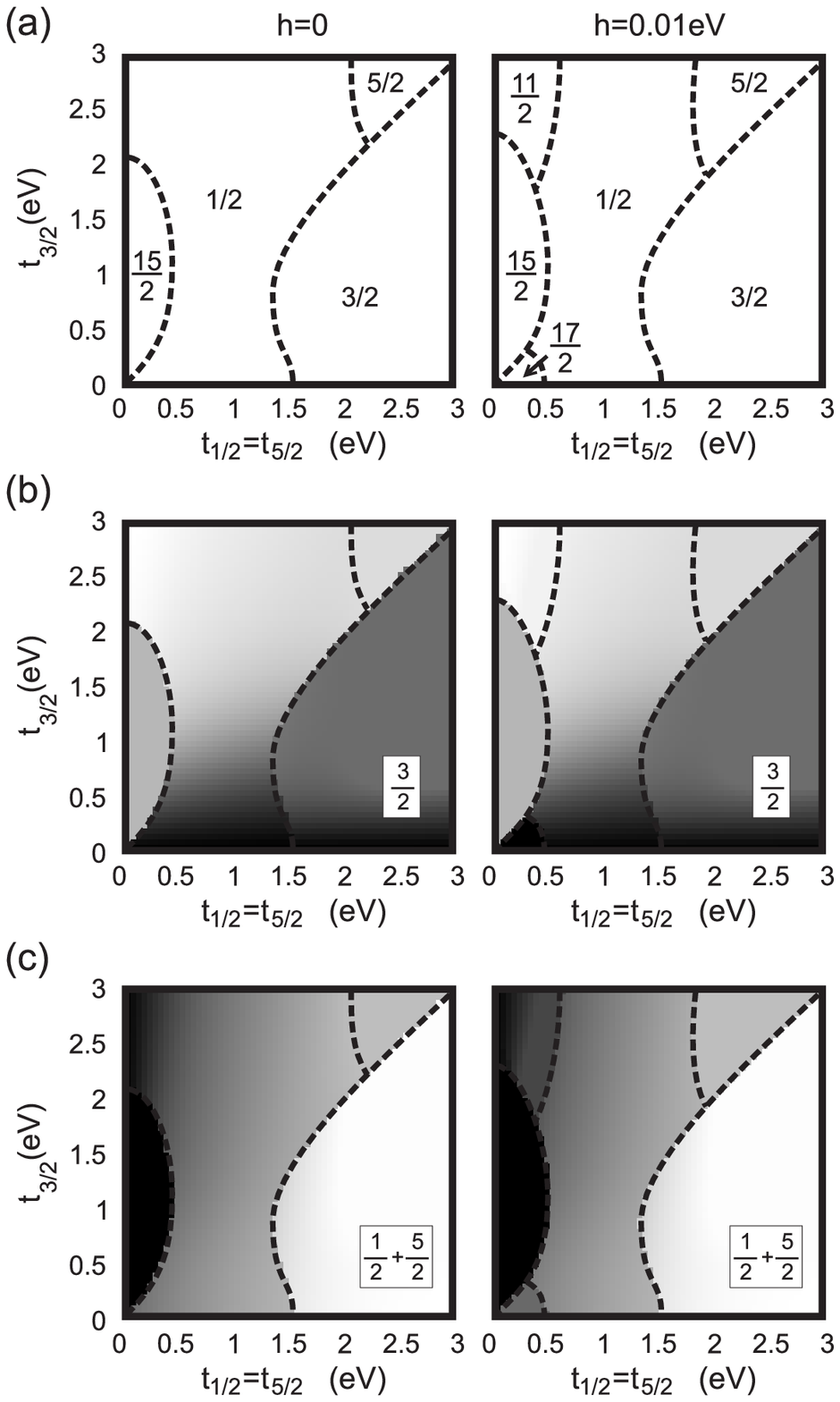}
\end{center}
\caption{
  Phase diagram for the Gutzwiller wavefunction in the  
  $t_{1/2}$\,-\,$t_{3/2}$ plane (with $t_{1/2}=t_{5/2}$) for vanishing 
  (left) and finite (right) ferromagnetic field.
  (a) Phases are characterized by their ${\cal J}_z$ expectation value.
  (b) Ratio  $T_{3/2}/t_{3/2}$  
  as gray-scale plots (black: 0, white: 1.4) and 
  (c) the sum of the corresponding ratios for $j_z$ = 1/2 and 5/2
  (black: 0, white: 2.4). 
  In comparison with Fig.~\ref{fig:LDA+U}, note the different plot ranges.
}\label{figGutz}
\end{figure}
}%%%%%%%%%%%%%%%%%%%%%%%%%%%%%%%%%%%%%%%%%%%%%%%%%%%%%%%%%%%%%%%%%%%%%%%%
\def\bildI{%%%%%%%%%%%%%%%%%%%%%%%%%%%%%%%%%%%%%%%%%%%%%%%%%%%%%%%%%%%%%%
\begin{figure}[t]
\begin{center}
  \includegraphics[width=0.85\columnwidth]{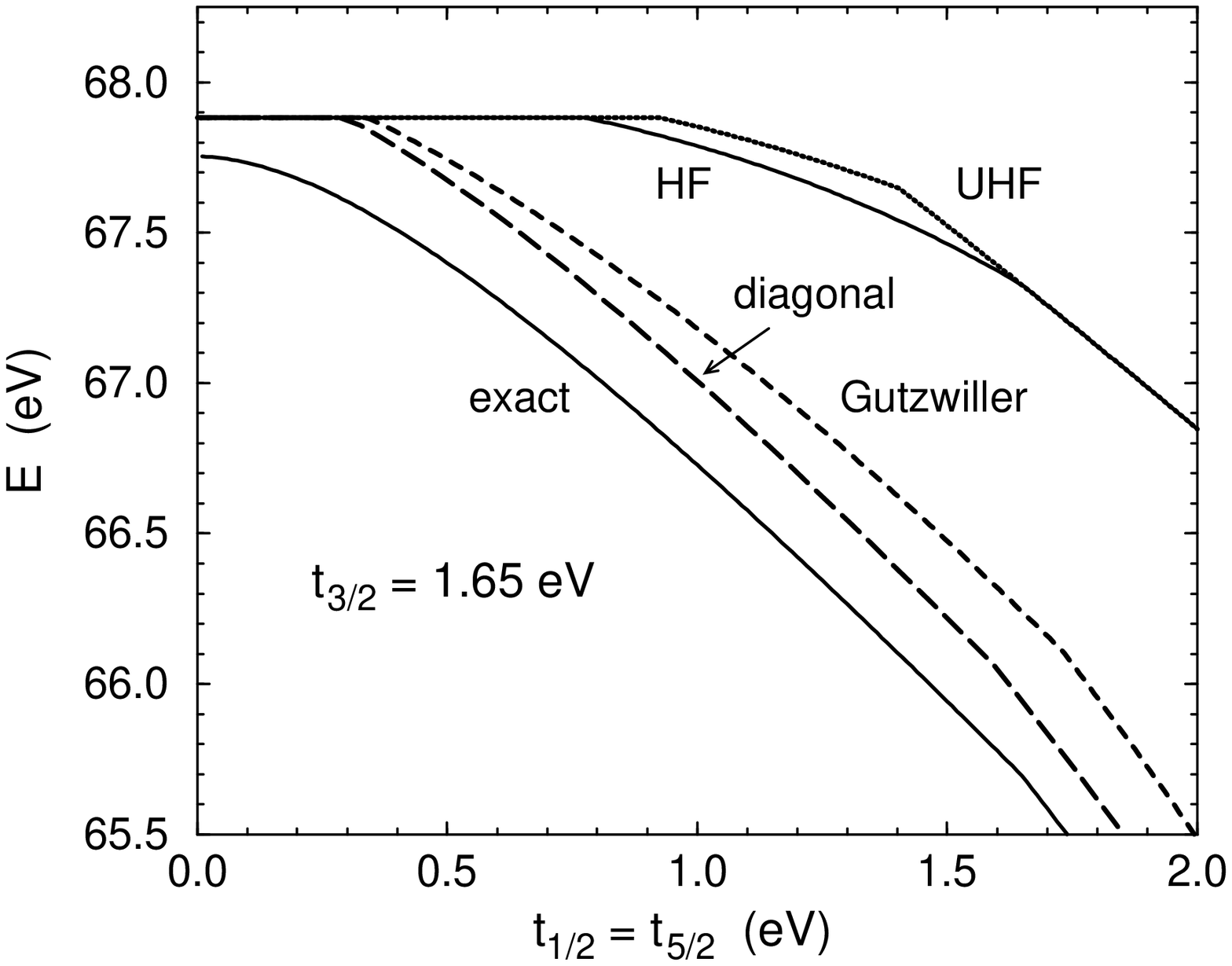}
\end{center}
\caption{
  Total energies for various approximation as function of 
  $t_{1/2}= t_{5/2}$ at fixed $t_{3/2}=1.65$\,eV:
  (exact) numerical diagonalization of the $600\times 600$ matrix, 
  see Fig.~\protect\ref{figCompare};
  (diagonal) diagonalization keeping only the diagonal interaction 
  matrix elements, see Fig.~\protect\ref{fig:LDA+U};
  (HF) Hartree Fock approximation, see Fig.~\protect\ref{figHFMTT};
  (UHF) Unrestricted Hartree Fock approximation, see Fig.~\protect\ref{figUHF};
  (Gutzwiller) variational wavefunction (\ref{eqGutzWave}), 
  see Fig.~\protect\ref{figGutz}.
}\label{figEnergies}
\end{figure}
}%%%%%%%%%%%%%%%%%%%%%%%%%%%%%%%%%%%%%%%%%%%%%%%%%%%%%%%%%%%%%%%%%%%%%%%%

\begin{document}
\title{Approximative treatment of $5f$-systems with partial localization 
due to intra-atomic correlations}
\author{E.~Runge and P.~Fulde}
\affiliation{Max-Plack-Institut f\"ur Physik komplexer Systeme, N\"othnitzer-Str.~38, 01187 Dresden, Germany}
\author{D.~V.~Efremov}
\altaffiliation[Present address: ]{%
Inst.\ Theoretische Physik, Technische Universit\"at, 01062 Dresden, Germany}
\affiliation{University of Groningen, Nijenborgh 4, 9747 AG Groningen, The Netherlands}
\author{N.~Hasselmann}
\affiliation{Institut f\"ur Physik, Johann-Wolfgang-Goethe Universit\"at, 60054 Frankfurt am Main, Germany}
\author{G. Zwicknagl}
\affiliation{Institut f\"ur Mathematische Physik, Technische Universit\"at Braunschweig, Mendelssohnstr.~3, 38106 Braunschweig, Germany}
\date{\today}
\begin{abstract}
Increasing experimental and theoretical evidence 
points towards a dual nature of the 5$f$ electrons 
in actinide-based strongly correlated
metallic compounds, with some 5$f$ electrons 
being localized and others delocalized. 
In a recent paper (PRB xxx, 2004),  
we suggested the interplay of intra-atomic correlations as 
described by Hund's rules and a weakly anisotropic 
hopping (hybridization) as a possible mechanism.
The purpose of the present work is to provide
a first step towards a microscopic description
of partial localization in solids by analyzing 
how well various approximation schemes perform 
when applied to small clusters.
It is found that  many aspects of 
partial localization are described appropriately
both by a variational wavefunction 
of Gutzwiller type and by a treatment which 
keeps only those interactions which are present in
LDA+U calculations.
In contrast, the energies and phase diagram 
calculated within the Hartree Fock approximation show little
resemblence with the exact results. 
Enhancement of hopping anisotropy by 
Hund's rule correlations are found in all 
approximations.
\end{abstract}
% Fermi-liquid theory and other phenomenological models}
\maketitle

\section{Introduction}

Many fundamental phenomena in solid-states physics can be
considered as resulting from the competition of the 
repulsive Coulomb interaction trying to keep electrons 
as far from each others as possible and the kinetic energy
term trying to delocalize electrons as much as possible. 
One of the most prominent examples
is the Mott-Hubbard transition from a 
narrow-band metal with delocalized electrons
to an insulator with  localized electrons.
Another one is band magnetism where the 
occupation of majority and minority bands differ
dramatically. 
The resulting magnetic moment is typically 
aligned along certain crystallographic axes 
by an anisotropy energy which is much weaker than 
typical electronic energies given by 
the bandwidths.
The recently suggested scenario of a dual 
nature of the 5$f$ electrons in some 
uranium-based heavy-fermion compounds 
combines features of both these examples: 
There is experimental\cite{Schoenes,Takahashi,Metoki,Bernhoeft,Yaouanc,Dressel02,Sato01}
and theoretical evidence\cite{Sato01,GYF02,Petit02,Petit03,GYF03,GEZJapan,YAF03,I} 
that in these materials some of the 5$f$ electrons with $j=5/2$ are   
localized and others are not. 
Which of the 5$f$~orbitals form local moments 
and which form bands will depend on 
the geometry of the crystallographic
environment of the particular actinide site, 
which leads to small differences of the hopping (hybridization) 
matrix elements.\cite{Hotta}
Band structure calculations augmented 
by a mass-renormalization of the
'free' 5$f$~electrons by crystal field excitations
of the 'localized' 5$f$~electrons
lead to very good agreement of both
areas and masses with dHvA data 
in UPt$_3$ and UPd$_2$Al$_3$.\cite{GYF02,GYF03,GEZJapan}
In both cases, the best agreement was achieved treating
the $j_z=\pm 3/2$ orbital as itinerant and including two
$f$~electrons in the ionic core.
Calculations within the so-called LDA+U scheme\cite{LDAU}
for UPd$_{3-x}$Pt$_x$ also suggest two fully occupied $f$~electrons
per U atom, however with $j_z=+5/2$ and $j_z=+3/2$
having occupations very close to unity.\cite{YAF03}
A treatment within the self-interaction corrected 
local density approximation\cite{SICLDA} (SIC-LDA) 
found  for  UPt$_3$ that the configurations of 
lowest energies have a local $f^{\,1}$ moment.\cite{Petit02}
The same authors found for UPd$_2$Al$_3$ 
a groundstate characterized by the coexistence 
of localized $f^{\,2}$ and delocalized Uranium $f$ electrons.
Rare-earth metals and their sulfides, 
i.e., 4$f$ electron systems, 
were also studied within the SIC-LDA and 
discussed in terms of localized and delocalized orbitals.\cite{Strange} 
However, the physical reasoning is quite different from the present one.  

In a recent paper,\cite{I}
we suggested the interplay of the intra-atomic 
correlations described by Hund's rules and a weakly 
anisotropic hopping as a possible 
mechanism leading to a separation of the 5$f$ orbitals 
into band-forming and local-moment-forming ones. 
The underlying idea can be described as follows. 
According to Hund's rules, atoms with three or two 
$f$~electrons have angular momenta of $J=9/2$ with 
10-fold degeneracy and $J=4$ with 9-fold degeneracy,
respectively. Consider now transitions
$M(j_z) = \langle 4J_z'|c_{j_z}| \frac{9}{2} J_z\rangle$   
between any given pair of those degenerate states
due to an annihilation operator $c_{j_z}$ 
stemming from the kinetic energy term of the Hamiltonian. 
Obviously, these matrix elements depend strongly on $j_z$, 
leading to much larger transition amplitudes 
for some $j_z$ than for others. 
Turning this argument around, the following scenario 
evolves: 
The largest product of the matrix elements $M(j_z)$ multiplied 
by the overlap $t_{j_z}$ between the corresponding 
atomic wavefunctions of neighboring sites determine
which pairs of $J=9/2$ and $J=4$ configurations
are realized, thereby maximizing the corresponding 
energy gain. 
In most cases, the other hopping processes are then surpressed. 
This scenario is supported by exact diagonalizations 
of small clusters presented in 
Refs.~\onlinecite{I} and~\onlinecite{N4}.

This mechanism differs from other routes to 
partial localization and a dual nature 
in strongly correlated systems with degenerate $d$ or $f$ orbitals
which were suggested, e.g., for manganites. 
For the latter, the physics is different: 
The degeneracy is split by strong crystal field 
with the lower $t_{2g}$ orbitals forming a 
localized high-spin state 
while the $e_g$ electrons are delocalized.

The numerical effort of exact cluster diagonalizations 
grows rapidly with system size. 
Any quantitative description of bulk material 
will necessarily involve some approximations. 
Over the last decades, considerable experience has been 
accumulated about the capability of various approximation 
schemes to describe many-body phenomena such as 
the above mentioned Mott-Hubbard transition and band 
magnetism.\cite{Gebhard} 
The purpose of the present paper is to provide
a first step towards a microscopic description
of partial localization in solids by analyzing how various 
approximation schemes perform when applied to 
the cluster model of Ref.~\onlinecite{I}.
The latter is briefly summarized in Sect.~\ref{secModel}.
In Sect.~\ref{secLDAU} and \ref{secHF}, 
two widely used approximation schemes are applied. 
In the first, the Coulomb interaction is approximated 
by its diagonal part (direct term plus exchange term) in
analogy with a modified local density approximation
to density functional theory (LDA+U). 
The second scheme is mean-field theory, also known as 
the Hartree-Fock approximation (HF).
In Sect.~\ref{secGutzwiller}, we study a
variational wavefunction similar to that 
proposed by Gutzwiller in his famous
paper on the ferromagnetism of 
transition metals.\cite{Gutzwiller}
We end in Sect.~\ref{secConclusion}
with a comparision of the various approximative treatments
and with some tentative conclusions and 
speculations about the performance of these
approximation schemes in bulk materials. 

\section{The model Hamiltonian\label{secModel}}

The simplest model describing the interplay between intra-atomic
correlations of $f$~electrons and weakly anisotropic hopping 
is considered in Ref.~\onlinecite{I}:
\begin{eqnarray}
  H & = & H_{\mathrm{Coul}} + H_{t} + H_{h} 
  \,.
  \label{eqHHHH}
\end{eqnarray}
The Coulomb interaction is written in terms of 
the usual fermionic operators $c_{j_z}^+(n)$ 
creating an electron on site $n$ with  angular momentum $j=5/2$ 
and $z$ component $j_z$ as  
\begin{eqnarray}
  \label{eqHCB}
  \lefteqn{\hspace*{-2mm}H_{\mathrm{Coul}} =}  \\
  && \hspace*{-6mm}\frac{1}{2}
  \sum_{n} \sum_{j_{z1},\dots,j_{z4}} 
  U_{j_{z_1}j_{z_2}j_{z_3}j_{z_4}}
  c_{j_{z1}}^         {+} (n)c_{j_{z2}}^         {+} (n)
  c_{j_{z4}}^{\phantom{+}}(n)c_{j_{z3}}^{\phantom{+}}(n)
   \nonumber \,,  
\end{eqnarray}
with Clebsch Gordon coefficients $C_{\dots}^{\dots}$ and   
Coulomb parameters $U_J$
\begin{eqnarray}
  U_{j_{z_1}j_{z_2}j_{z_3}j_{z_4}} &=& 
  \Big\langle j_{z_1}j_{z_2} | \hat U | j_{z_3}j_{z_4}  \Big\rangle 
   \label{eqHCA} \\
  &=& 
  \sum_J U_{J}  \,
  C^{JJ_{z}}_{5/2,j_{z1}\,;\,5/2,j_{z2}}
  C^{JJ_{z}}_{5/2,j_{z3}\,;\,5/2,j_{z4}}
  \nonumber
  \,.
\end{eqnarray}
The kinetic energy operator describes 
hopping (hybridization) between all pairs of neighboring sites $\langle nm\rangle$
\begin{eqnarray}
  H_t & = & - \sum _{\langle nm\rangle, j_{z}}\, t_{j_{z}}\Big( 
    c_{j_{z}}^{+}(n)\,c_{j_{z}}^{\phantom{+}}(m)+h.c.\Big)
  \,.
  \label{eqHT}
\end{eqnarray}
A ferromagnetic or antiferromagnetic 
external or internal field
\begin{eqnarray}
  H_h & = &  -\sum _{n,j_{z}}\, h_{n}\,\, j_{z}\,\,c_{j_{z}}^{+}(n)
  \,c_{j_{z}}^{\phantom{+}}(n)
  \,
  \label{eqHh}
\end{eqnarray}
is included as well.

For the cases of two sites occupied by 5 electrons and for three sites
and 8 electrons, iterative sparse-matrix 
diagonalization easily yields the respective ground states
and some low-lying eigenstates.
The sizes of the Hamilton matrices 
are $600\times 600$
and $18,000\times 18,000$, 
respectively, provided that site configurations are
limited to $f^{\,2}$ and $f^{\,3}$ states, 
and $792\times 792$ and $61,758\times 61,758$
if this restriction is lifted.\cite{N4}
For the sake of a transparent presentation, we focus in the 
present paper on the two-site model and stay 
in the 600~dimensional $f^{2}$-$f^{3}$ subspace. 

The $U_J$ parameters are more robust to changes in
the chemical composition or the application of pressure
than the the $t_{j_z}$. Thus, 
we use  $U_J$ parameters appropriate for UPt$_3$,
$U_{J=4}=17.21 eV$, $U_{J=2}=18.28 eV$, $U_{J=0}=21.00 eV$,
\cite{GYF02}, and study the phase diagram in the
$t_{j_z}$ space.  
As in Ref.~\onlinecite{I}, we choose $t_{1/2}=t_{5/2}$
in the figures. The phase diagrams of the 
model Hamiltonian~(\ref{eqHHHH})-(\ref{eqHT}) 
obtained by exact numerical diagonalization
is reproduced in Fig.~\ref{figCompare}.
\bildA%%%%%%%%%%%%%%%%%%%%%%%%%%%%%%%%%%%%%%%%%%%%%%%%%%%%%

\section{Diagonal interaction matrix elements\label{secLDAU}}

Almost all electronic structure calculations for extended systems and
large molecules are performed either in the framework of the 
local density approximation (LDA) or using the Hartree-Fock approximation.
The former accounts for correlations only in a rather crude way,
while the latter neglects them altogether. 
Various recent papers suggested the inclusion of 
strong local correlations via an on-site energy-term
of Hubbard type, $\sim U \hat{n}_j \hat{n}_k$,  
in LDA calculations within the LDA+U scheme\cite{LDAU} 
or the related  SIC-LDA.\cite{Petit02,Petit03,Strange}
This is a promising 
route for some groundstate properties of highly correlated systems. 

The key approximations of the LDA+U scheme are on the one hand the 
LDA itself and on the other hand the inclusion of
only the diagonal interaction matrix elements 
(direct and exchange)
\begin{equation}
\label{eqLDAPlusU}
\tilde{U}_{j_{z_1}j_{z_2}j_{z_3}j_{z_4}} \! = (\delta_{j_{z_1}j_{z_3} }
\delta_{j_{z_2}j_{z_4} } \!+\delta_{j_{z_1} j_{z_4}}
\delta_{j_{z_2}j_{z_3}} ) \,U_{j_{z_1}j_{z_2},j_{z_3}j_{{z_4}_{\vphantom{g}}}}
  .
\end{equation}
The fact that this approximation corresponds to the 
LDA+U approach is best seen by inserting it into the 
Coulomb Hamiltonian~(\ref{eqHCB}) 
and rewriting the latter in 
terms of $\hat{n} _{a,j} =  c^{+}_{a,j} c^{\phantom{+}} _{a,j}$
as 
\begin{equation}
\label{eqLDAPlusU2}
  \tilde{H}_{\mathrm{Coul}}= \sum_{a=1,2}\ \sum_{j_zj_z'} 
  \big\langle j_{z}j_{z}' \big| \hat U\big|j_{z}j_{z}'\big\rangle 
  \,\, \hat n _{aj_{z}} \hat n _{aj_{z}'}  
  \,.
\end{equation}
Henceforth, we will refer to Eq.~(\ref{eqLDAPlusU2}) as 
`diagonal approximation.'

\bildB%%%%%%%%%%%%%%%%%%%%%%%%%%%%%%%%%%%%%%%%%%%%%%%%%%%%%
Figure~\ref{fig:LDA+U} presents  the resulting phase diagrams  
in the plane of hopping parameters $t_{1/2}=t_{5/2}$ 
and $t_{3/2}$.
A visually striking difference to the result of the
full calculation, Fig.~\ref{figCompare}, 
is the complete absence of the isotropic line 
$t_{1/2}=t_{3/2}=t_{5/2}$ in the phase diagram
of the diagonal approximation to the Hamiltonian.
This reflects the fact that the 
condition~(\ref{eqLDAPlusU}) of 
diagonality of $\tilde U$ with respect to $z$-component 
indices breaks rotational invariance.
The $z$ component of the total angular momentum
remains a good quantum number. 
We note that the diagonal approximation overemphasizes
the parameter space of phases with either 
strong ferromagnetic  or strong  antiferromagnetic 
alignment along $z$, i.e., ${\cal J}_z=15/2$ and ${\cal J}_z=1/2$.
The simplified treatment quite correctly predicts the 
ferromagnetic `high-spin' state  ${\cal J}_z=11/2$ for 
$t_{3/2} \gg t_{1/2}=t_{5/2}$.
In general, the approximation fares worst in the case 
of small hopping anisotropy.
The ${\cal J}_z=5/2$ phase is missing, while a ${\cal J}_z=3/2$ state 
is predicted only for rather strong hopping. 

\section{Hartree-Fock approximation\label{secHF}}

The Hartree-Fock approximation is one of the most widely 
used calculational approaches for interacting systems.   
For its  implementation, 
one uses the single-particle density matrix 
\begin{equation}
 \label{eqHFDensityMat}
 n_{aj_z,a'j_z'}  =  \sum_{\nu=1}^{N_{\mathrm{tot}}} 
 {\varphi^{(\nu)} _{aj_z} }^* \varphi^{(\nu)}_{a'j_z'} 
 \,,
\end{equation}
written as sum of $N_{\mathrm{tot}}$  HF orbitals, 
which are superpositions of  atomic orbitals 
described by the $c^{+}_{j_z}$
\begin{eqnarray}
 \label{eqHFAllgemein}
 | \varphi^{(\nu)} \rangle &=&  \sum_{a=1,2} \sum_{j_z=-5/2}^{5/2} 
 \varphi^{(\nu)}_{aj_z} \,\, c^+_{j_z} \,\, |0\rangle 
 \,.
\end{eqnarray}
We impose no restrictions on the single-particle
states, which are arbitrary combinations of angular momenta.   
The system of HF equations is\cite{GrossRunge} 
\begin{equation}
 E_\nu \,\, \varphi^{(\nu)}_{aj} =  t_j \,\, \varphi^{(\nu)}_{\overline{a}j} +
  2\,\, \sum_{j'} v^{(HF)}_{a,jj'}  \varphi^{(\nu)}_{aj'}
  \,,
\end{equation}
with
\begin{eqnarray}
 \label{eqHFPotential}
 v^{(HF)}_{a,jj'} &=& \sum_{kk'} \,n_{ak,ak'}\, 
 \langle jk |\hat U | j'k'\rangle 
\end{eqnarray}
and $a, \overline a$ denoting the two different sites.
The two-particle matrix element
$ \langle jk |\hat U | j'k'\rangle $ 
comprises both the direct Hartree and the Fock contributions.
Note that it is not restricted to 
two pairwise equal indices in the $j_z$ basis. 
The total energy is obtained from
\begin{eqnarray}
 E_{\mathrm{total}} &=&  \sum_\nu E_\nu -  \frac{1}{2} \sum_\nu \sum_{jj'} 
     {\varphi^{(\nu)}_{j}}^*   v^{(HF)}_{a,jj'}    \varphi^{(\nu)}_{j'}
  \,,
\end{eqnarray}
where double-counting of the interaction energy is avoided 
in the usual way.\cite{GrossRunge}

In general, the HF solutions do not have the total ${\cal J}_z$  
as a good quantum number. Therefore we cannot always use 
different ${\cal J}_z$'s to classify different phases. 
However, one still finds discontinuous jumps 
either of the expectation value $\langle {\cal J}_z \rangle $ or of its
derivatives with respect to $t_{j_z}$ or $h$ 
and we use these to define phase boundaries. 
Moreover
for some phases with small kinetic energy, 
self-consistent solutions 
$\varphi^{(\nu)}$ with good quantum numbers 
$j_z$ are found and, thus, ${\cal J}_z$ is 
a good quantum number in those cases. 
This applies to the phases with ${\cal J}_z=15/2$ and $17/2$.
For larger hopping parameters,
one finds  in general continuous variations of 
$\langle {\cal J}_z \rangle$.
We denote phases as `AM' or `FM' indicating antiferromagnetic
or ferromagnetic spin correlations, respectively. 
Half-integer-valued subscripts give the approximate $\langle {\cal J}_z \rangle $ 
expectation value found for $t_{j_z}$ parameters 
not too close to the phase boundaries. 
\bildC%%%%%%%%%%%%%%%%%%%%%%%%%%%%%%%%%%%%%%%%%%%%%%%% 

Overall, the Hartree-Fock  $t_{1/2}$\,-\,$t_{3/2}$  
phase diagram in Fig.~\ref{figHFMTT} shows only 
weak resemblance with the one from the exact calculation. 
The isotropic line is a phase boundary at zero field.

Figure~\ref{fig:spinSchemeMagnet} illustrates the occupation pattern of 
four important phases occuring in the phase diagrams shown in 
Figs.~\ref{figCompare}-\ref{figHFMTT}
and \ref{figUHF} of the present work  
as well as Fig.~5 of Ref.~\onlinecite{I}.
A large angular momentum as implied by Hund's rules is obtained 
if electrons occupy preferrentially states with either large 
positive or large negative angular momentum projections $j_z$
-- of course subject to the Pauli principle. 
Occupation of states with large positive or negative 
angular momentum projections 
is favored also by local (molecular) magnetic fields. 
In particular, the antisymmetrized product states
$\big|\pm \frac{5}{2},\pm\frac{3}{2},\pm\frac{1}{2}\big\rangle $
$\times$ $\big|\pm \frac{5}{2},\pm\frac{3}{2}\big\rangle $
mi\-ni\-mize simultaneaously
the Coulomb energy and the magnetic energy. 
For not too strong hopping, the two sites \mbox{(anti-)}\,align
their angular momenta in a (anti-)\,ferromagnetic field
configuration, see left (right) column of Fig.~\ref{fig:spinSchemeMagnet}.
\bildD%%%%%%%%%%%%%%%%%%%%%%%%%%%%%%%%%%%%%%%%%%%%%%%%%%%%%%%%%%%%%%%%%%%%

Two crucial differences between FM and AF fields are obvious:
(\textit{i}) As drawn, in the FM case both sites have an $f$~occupation
$n_f=2.5$, whereas  in the AM case 
a small charge disproportionation $\Delta n_f < 1$ 
can be present.
The AM configurations of Fig.~\ref{fig:spinSchemeMagnet}
are, of course, degenerate due to the symmetry
with respect to a simultaneous exchange of 
site index and $j_z\leftrightarrow -j_z$.

(\textit{ii}) The FM and AM cases differ with respect to a
possible quantum phase transition when going
from $t_{ 3/2} <  t_{ 1/2}$ 
to   $t_{ 3/2} >  t_{ 1/2}$
(compare in Fig.~\ref{fig:spinSchemeMagnet} the  
upper row with the lower row). 
In the AF case, with increasing 
$t_{1/2}$ the $j_z=1/2$~electron starts to fluctuate 
more between the sites,
whereas the $j_z=3/2$~electron
fluctuates less. This is a continuous cross-over,
which however takes place rather suddenly because 
$j_z=1/2$~fluctuations 
imply charge fluctuations which then suppress 
$j_z=3/2$~fluctuations, and vice versa.
In a level diagram such as the inset 
of Fig.~4 in Ref.~\onlinecite{I}, this corresponds 
to an avoided crossing.

In contrast, in the FM case one of 
 $j_z=1/2$ or $j_z=3/2$ is not fully occupied. 
The transition is not a cross-over, but a simple 
level crossing with discontinuous change 
of ${\cal J}_z$ and orbital occupation.
This difference between sharp discontinuities on 
the FM side and smooth cross-overs on the AF side will be
seen repeatedly in this work.
We will see below that within the Hartree-Fock approximation 
the states $FM_{15/2}$, $FM_{17/2}$, and  $AM_{1/2}$ 
dominate indeed  in the phase
diagram for finite fields and not too large hopping. 

One further cautionary remark is in place: 
The representation in Fig.~\ref{fig:spinSchemeMagnet}
shows only the occupation pattern,
but does not reflect the correlations between different 
configurations.
     
Strongly anisotropic hopping $t_{3/2} \gg t_{1/2}$ 
lead in the full solution to FM$_{15/2}$, 
i.e.\ to ferromagnetic correlations 
and single occupancy of the binding $j_z=+\frac{3}{2}$
state. The corresponding wavefunction  
\begin{equation}
|\Psi \rangle_{\frac{15}{2}} = 
\frac{c^{+}_{3/2}(a)+c^{+}_{3/2}(b) }{\sqrt{2}}\,
 c^{+}_{5/2}(a)
 c^{+}_{1/2}(a)
 c^{+}_{5/2}(b)
 c^{+}_{1/2}(b)\, | 0 \rangle
 \label{eq:FerroHighState15}
\end{equation} 
is a simple Slater determinant.
Thus, it is not surprising that HF works well 
for this case.
The parameter space of the FM$_{15/2}$ phase is, however, 
much larger in the HF approximation than 
in the exact calculation. 
It shrinks with increasing antiferromagnetic field 
and increases with increasing ferromagnetic field.

Outside the ferromagnetic phases in Fig.~\ref{figHFMTT}, 
the correlations are antiferromagnetic and 
$\langle {\cal J}_z \rangle \lesssim 3/2$.
A kink separates a phase with 
$\langle {\cal J}_z \rangle \approx 1/2$, 
visualized in Fig.~\ref{fig:spinSchemeMagnet} as AM$_{1/2}$, 
from a phase with 
$\langle {\cal J}_z \rangle \approx 1$  at or near the isotropic line
and $\langle {\cal J}_z \rangle \sim 3/2$ at larger values of $t_{3/2}$
which is denoted by AM$_{3/2}$ in Fig.~\ref{fig:spinSchemeExtra}.
In a small bubble bordering the
ferromagnetic FM$_{15/2}$ phase, a phase exists 
which we denote by AM$^{\mathrm{S}}_{3/2}$, see Fig.~\ref{fig:spinSchemeExtra}.
It can be considered as the symmetrized version of AM$_{3/2}$.
The binding linear combinations with 
$j_z=+3/2$ and $j_z=-3/2$
are both half occupied and contribute equally to 
the kinetic energy gain by hopping (hybridization).
Strongly anisotropic hopping with $t_{1/2} \gg t_{3/2}$
leads to AM$_{1/2}$, i.e.\ to an antiferromagnetic
arrangement with broken left-right symmetry,
i.e., $0<\Delta n_f \lesssim 1$. 

A region of maximal ferromagnetic moment (FM$_{17/2}$) is
found for a ferromagnetic field which is 
still weak compared to the bare hopping strengths. 
For larger fields, the ferromagnetic  phase FM$_{5/2}$
with `intermediate' spin is found close to
isotropic line. 
In this phase, all five electrons 
contribute to the kinetic energy gain.

Figure~\ref{fig:spinSchemeExtra} shows 
also a phase denoted by FM$_{11/2}$, 
which is of overall ferromagnetic nature
but dominated by $t_{3/2}$ hopping. 
It appears in Fig.~\ref{figCompare} and \ref{fig:LDA+U}
as well as in Fig.~\ref{figGutz} below.
\bildE%%%%%%%%%%%%%%%%%%%%%%%%%%%%%%%%%%%%%%%%%%%%%%%%%%%%%%%%%%%%%%%%%%%

\subsection{Partial localization\label{secHFPartial}}

In order to quantify the degree of localization 
of a given $j_z$ orbital by local correlations,
we study the ratio of the
$j_z$-projected kinetic energy $T_{j_z}$  
and the bare matrix element $t_{j_z}$\cite{I}
\begin{equation} 
  \label{eqDefDeltaTJ}
  \frac{T_{j_z}}{t_{j_z}} = 
  \sum_{\langle nm \rangle, \pm} \langle\Psi_{\mathrm gs} | 
  (  c^{+}_{\pm j_z}(n)\,  c_{\pm j_z}^{\phantom{+}}(m) + h.c. ) 
|\Psi_{\mathrm gs} \rangle 
  \, ,
\end{equation} 
with $-2\leq T_{j_z} / t_{j_z}\leq +2$ 
for two sites.
A small ratio $T_{j_z}/t_{j_z}$
indicates partial suppression of hopping for electrons in the
$\pm j_z$ orbitals.
For all cases considered in Ref.~\onlinecite{I}, 
we found small ratios of $T_{j_z}/t_{j_z}$ 
for the smaller $t_{j_z}$, 
and values close to one or two for the dominant
hopping matrix element. 

In Fig.~\ref{figHFPartial}, we study the same quantity 
for the Hartree-Fock solutions. 
Again if $t_{3/2}\gg t_{1/2},t_{5/2}$, 
ratios close to one are found for $T_{3/2}/t_{3/2}$,
but much smaller values for $(T_{1/2}+T_{5/2})/t_{1/2}$,
and vice versa. 
At first glance, this may look surprising since the
Hartree-Fock method can not describe the correlations 
underlying the dual nature of the $f$~electron in our 
model system.    
A closer look reveals that the correlations are 
simulated by symmetry breaking, i.e.\ 
by driving the corresponding 
occupations close to zero or two. 
In contrast, the full solution shows a reduced 
hopping also for intermediately occupied orbitals.
We conclude that HF mimics correlations by 
overemphasizing occupation differences.
Thus, further studies are needed to gauge the
usefulness of HF for bulk systems with partially 
localized 5$f$ electrons.  
\bildF%%%%%%%%%%%%%%%%%%%%%%%%%%%%%%%%%%%%%%%%%%%%%%%%%%%%%%%%%%%%%%%%%%%%%%%%%%%%%

\subsection{Hartree-Fock with $j_z$ eigenstates (UHF)\label{secUHF}}
If the variational HF space is reduced further by requiring 
$j_z$-diagonal density matrices and self-consistent 
potentials 
\begin{eqnarray}
  \label{eqUHFPotential}
  v^{(UHF)}_{a,j} &=& 
  \sum_{k} n_{a,k}\,\, \langle jk |\hat V | jk\rangle 
  \,,
\end{eqnarray}
the $z$~axis plays a special role (quantization axes), 
as it did in Sect.~\ref{secLDAU}, and the spherical symmetry is broken.
Following the usual nomenclature, 
we refer to this approximation which allows different 
orbitals for different angular momenta but
prescribes angular momenta as orbital quantum number, as 
the unrestricted Hartree-Fock (UHF)\cite{GrossRunge}
-- even though the wavefunction space is restricted in comparison to
Eq.~(\ref{eqHFAllgemein}).
The resulting phase diagrams, Fig.~\ref{figUHF},
are not identical, but close to those
of the full HF solution in Fig.~\ref{figHFMTT}.
Not surprisingly, the isotropic line does no longer
play a special role. The corresponding phase
boundary has moved to higher $t_{3/2}$ values, 
a consequence of the variational restriction favoring the maximally 
antiferromagnetic alignment with ${\cal J}_z=1/2$. 

Note that ${\cal J}_z$ is now a good quantum number. In particular, the 
border between AM$_{3/2}$ and AM$_{1/2}$ is here
easily identifiable, the kink being  replaced by a jump. 
A noticeable difference between the HF and the UHF 
phase diagram is that AM$^{\mathrm{S}}_{3/2}$, 
i.e.\ the symmetrized version of $AM_{3/2}$, 
can not be found within the UHF approximation, 
because all six $j_z$ values are partially occupied
in this phase, cf.\ Fig.~\ref{fig:spinSchemeExtra}. 

As an aside, we remark that numerical convergence 
in solving the restricted equation (\ref{eqUHFPotential}) 
is obtained in a small fraction of the time needed
for solving the HF equations (\ref{eqHFDensityMat}),
(\ref{eqHFPotential}).
\bildG%%%%%%%%%%%%%%%%%%%%%%%%%%%%%%%%%%%%%%%%%%%%%%%%%%%%%%%%%%%%%%%%%%%%%%%%

\subsection{Charge disproportionation}
The Hartree-Fock phase diagram is surprisingly rich with several 
phases showing charge disproportionation, i.e.\ a finite
expectation value $0<\Delta n_f <1$ 
of the difference of site occupations.     
Note in particular the phase boundary in Fig.~\ref{figHFMTT}
between the closely related phases AM$_{3/2}$ and AM$^{\mathrm{S}}_{3/2}$ at 
vanishing or staggered field.

The exact groundstates of the small cluster does not show 
charge disproportionation, because a suitable superposition of 
the two symmetry-related configurations will lead to a
finite energy gain,
completely analogous to the absence of ferromagnetism
in small clusters at vanishing external fields. 
However, in both cases the energy gain can be extremely small
if the configurations are sufficiently different. 
This feature of the numerically exact small cluster 
solution will often not survive the thermodynamic limit.
In contrast, mean-field approaches easily 
lead to symmetry-broken solutions even for small clusters.  
This is so, because they reduce charge fluctuations and therefore simulate
the effect of electron correlations.
When symmetry breaking is experimentally observed in bulk samples
it is overemphasized in a Hartree-Fock treatment.
Thus, we take the trend of the HF solution towards
charge  disproportionation only as a hint 
that it might be worthwhile to look for, e.g., 
phase transitions driven by charge ordering in
systems with partially localized 5$f$ orbitals.
  
\section{Gutzwiller's wavefunction\label{secGutzwiller}}

The exact groundstates found in Ref.~\onlinecite{I}
often have a large overlap with a Gutzwiller-type
wavefunction\cite{Gutzwiller} of the form
\begin{equation}
   \big|\Psi_{G}\big\rangle = 
   \hat{\cal P}_{f^{2}f^{3}} \prod_{\nu=1}^5 \tilde{c}^{+}_\nu
   \big|0\big\rangle
   \,,
\end{equation}
where the projector $\hat{\cal P}_{f^{2}f^{3}}$ retains only
local $f^{\,2}$ and  $f^{\,3}$ configuration. 
The single-particle orbitals
generated by the $\tilde{c}^{+}_\nu$ are 
subject only to the orthogonality requirement. 
In  most actual applications of the Gutzwiller ansatz, 
simple trial wavefunction are used
for the single-particle states,   
e.g., the non-interacting, possibly polarized, Fermi sea.
Here, we use the ansatz that each orbital has $j_z$ as a 
`good quantum number,' i.e.
\begin{equation}
   \big|\Psi_{G}\big\rangle =  \hat{\cal P}_{f^{2}f^{3}} \prod_{\nu=1}^5 
   \Big( \cos \theta_\nu c^{+}_{j_z^{(\nu)}}(a) + 
         \sin \theta_\nu c^{+}_{j_z^{(\nu)}}(b) \Big ) 
   \big|0\big\rangle
   \,,
   \label{eqGutzWave}
\end{equation}
with variational parameters $\theta_\nu$.
Both, the numerator and the denominator of the energy expectation
value 
\begin{equation}
   E[\{j_z^{(\nu)}\},\{\theta_\nu\}] = \frac
   { \langle\Psi_{G}  \mid H \mid \Psi_{G} \rangle }
   { \langle\Psi_{G}     \mid     \Psi_{G} \rangle }
   \label{eqEnergy}
\end{equation}
are polynomial expression in $\cos \theta_\nu$ and 
$\sin \theta_\nu$. Note that in the case 
of double occupancy
$j_z^{(\nu_1)} = j_z^{(\nu_2)}$  
the corresponding variational parameters
$\theta_\nu$ drop out of Eq.~(\ref{eqEnergy}). 

We performed numerical minimizations 
with respect to both the occupation pattern $\{j_z^{(\nu)}\}$ 
and the set of parameters $\{\theta_\nu\}$.
In practice,  the expectation value~(\ref{eqEnergy})
is obtained explicitly from the 
Hamilton matrix in the many-particle 
site-spin representation (600$\times$600) 
in terms of  $\theta_\nu$. 
Then \textsc{Mathematica}'s implementation of 
Brent's algorithm is used for 1-dimensional, 
3-dim.,  and 5-dim.\ numerical 
minimization in the case of 2, 1 and 0 doubly 
occupied angular momenta $j_z$, respectively.
Even the five-dimensional minimum search is 
fast. Unfortunately, many occupation pattern
have to be tried and different random initial 
conditions yield different answers. 
This makes an additional loop over various 
initial conditions necessary.
\bildH%%%%%%%%%%%%%%%%%%%%%%%%%%%%%%%%%%%%%%%%%%%%%%%%%%%%%%%%%%%%%%%%%%%%%%%%%%%%%

The obtained phase diagram is reproduced in 
Fig.~\ref{figGutz}a. Again, 
${\cal J}_z$ is a good quantum number because 
the corresponding operator commutes with $\hat{\cal P}_{f^{2}f^{3}}$. 
Both panels show a transition from a strongly polarized 
${\cal J}_z=15/2$ phase to the state of lowest possible polarization, 
i.e., ${\cal J}_z = 1/2$. 
A small pocket with ${\cal J}_z=17/2$ is found at finite field. 
The phase diagram for small antiferromagnetic fields (not shown) 
is close to that for vanishing field with the 
high-spin phase ${\cal J}_z=15/2$ being slightly reduced.
As in the exact solution and in all other approximation
schemes, the energy varies continuously (not shown).

The phase diagram of the Gutwiller 
wavefunction~(\ref{eqGutzWave}) is remarkably close to  
the exact diagonalization of the Hamiltonian 
keeping only diagonal matrix elements (`LDA+U') 
in Sect.~\ref{secLDAU}. We will see below that this
is true for the total energies as well.
This should not come as an surprise, because 
both approaches effectively reduce charge fluctuations  
but do not include angular correlations equally well.

\subsection{Partial localization\label{secGutzPartial}}

In Fig.~\ref{figGutz}b and \ref{figGutz}c, we study the quantity
$T_{j_z}/t_{j_z}$ for the Gutzwiller type ansatz 
(\ref{eqGutzWave}). 
If either $t_{3/2}\gg t_{1/2}$ or  $t_{1/2}\gg t_{3/2}$, 
we find again a clear suppression of the non-dominant 
hopping. 
It is interesting that in the ${\cal J}_z=1/2$ phase
one can go from one case to the other without crossing a 
phase boundary. 
This is interesting in so far as 
a Gutzwiller wavefunction describes this important point  
without having to introduce phase transitions by symmetry breaking.

\section{Summary and Outlook\label{secConclusion}}

We have studied the interplay and competition  
of intra-atomic correlations and an anisotropic
kinetic energy operator
by applying three different approximative many-body 
treatments to the model of Ref.~\onlinecite{I}. 
None of them leads to fully satisfactory results.
But, the diagonal approximation inspired by the LDA+U approach,
i.e.\ keeping only the matrix elements~(\ref{eqLDAPlusU}),
gives a reasonable good phase diagram with 
most important phases of the exact solutions being 
present.
This finding should encourage further applications
of LDA+U and related approaches such as SIC-LDA 
for groundstate properties of actinide heavy-fermion 
materials. 
We mentioned already in the introduction that some
LDA+U calculations indeed show partial localization.
However, one can not expect to reproduce within these schemes
the small energy scale responsible for the 
heavy-fermion character of the low-energy excitations. 

The numerically easier UHF approximation with $j_z$-diagonal
orbitals agrees quite well with the result of the most general
HF ansatz.
However, both yield phase diagrams with only 
weak resemblance with the one from the exact calculation. 
The substantial suppression of the subdominant hopping
found in Fig.~\ref{figHFPartial} might be somehow fortituous.

These observations are confirmed by the total energies
given in Fig.~\ref{figEnergies} for all approximation
schemes along a line $t_{3/2}=1.65$\,eV in the 
$t_{1/2}$\,-\,$t_{3/2}$ plane.
The particular $t_{3/2}$ value was chosen such that 
a large number of phase boundaries is crossed.
Clearly, at some phase boundaries kinks of the total energy 
are seen (related to level crossings)
whereas  
at others higher derivatives change more or less discountinuously 
(avoided crossings).
\bildI%%%%%%%%%%%%%%%%%%%%%%%%%%%%%%%%%%%%%%%%%%%%%%%%%%%%%%%%%%%%

We note again that for the exact numerical diagonalization and 
for the calculation
keeping only state-diagonal interaction matrix elements 
we surpressed  configurations $f^n$ with $n\not = 2,3$, 
i.e., we worked in the 600 dimensional space rather than  
the 792 dimensional space. Giving up these restrictions
lowers the corresponding energies slightly (not shown). 

All curves, except that of the exact solution start with 
the same value which is constant for a certain range of 
$t_{1/2}$ and $t_{5/2}$ values. 
The corresponding wavefunction is the Slater determinant
$|\Psi \rangle_{15/2}$, Eq.~(\ref{eq:FerroHighState15}),
which completely surpresses $t_{1/2}$ and $t_{5/2}$ hopping.
In contrast, the exact ground state near
$t_{3/2}=1.65$\,eV, $t_{1/2}=t_{5/2}=0$\,eV 
is not $|\Psi \rangle_{15/2}$, but has intermediate 
angular momentum ${\cal J}_z=5/2$ and gains energy from all hopping 
processes. 
This explains the lower energy and its dependence
on all hopping matrix elements.

The solutions of the most general Hartree-Fock scheme (HF)
and the case where the single-particle orbitals were 
restricted to be diagonal with respect to $j_z$ (UHF) 
differ only in small parts of the parameter space. 

While the present model was developed with actinide-based
materials in mind, it seems worthwhile to examine
the role of intra-atomic correlations in other materials,
most noticeable in transition metal compounds.    

In conclusion, we find 
the solutions for the modified Hamiltonian containing 
the diagonal matrix elements only and the expectation value 
of the original Hamiltonian with generalized
Gutzwiller variational wave function (\ref{eqGutzWave})
are quite close, because 
both approaches effectively reduce charge fluctuations  
but do not include angular correlations equally well.
The subtle angular correlations which determine the magnetic 
character are accounted for only in limiting cases.

\section{Acknowledgments}
We thank Profs.\ P.~Kopietz, B.~Marston, W.~Nolting,
F.~Pollmann, and A.~Yaresko for stimulating discussions.
D.E.\ thanks the Netherlands Foundation for Fundamental 
Research (FOM) for financial support.

%\bildA\par\bildB\par\bildC\par\bildD\par\bildE\par
%\bildF\par\bildG\par\bildH\par\bildI

\end{document}